# RubberEdge: Reducing Clutching by Combining Position and Rate Control with Elastic Feedback

*Géry Casiez[†], Daniel Vogel[††], Qing Pan[†] and Christophe Chaillou[†]*

[†]LIFL & INRIA Futurs
University of Lille, FRANCE
{*gery.casiez|qing.pan|christophe.chaillou*}*@lifl.fr*

[††]Department of Computer Science
University of Toronto, CANADA
*dvogel@dgp.toronto.edu*

**ABSTRACT**

Position control devices enable precise selection, but significant clutching degrades performance. Clutching can be reduced with high control-display gain or pointer acceleration, but there are human and device limits. Elastic rate control eliminates clutching completely, but can make precise selection difficult. We show that hybrid position-rate control can outperform position control by 20% when there is significant clutching, even when using pointer acceleration. Unlike previous work, our RubberEdge technique eliminates trajectory and velocity discontinuities. We derive predictive models for position control with clutching and hybrid control, and present a prototype RubberEdge position-rate control device including initial user feedback.

**ACM Classification:** H5.2 [Information interfaces and presentation]: User Interfaces. - Graphical user interfaces

**INTRODUCTION**

For the most part, a relative position control device, such as the mouse, will perform better than a rate control device, such as a joystick [6,9]. However, a potential issue with position control devices is when *clutching* – the momentary recalibration to avoid running out of input area – becomes more frequent, taking additional time [12,16]. Recently the resolution of digital displays has increased significantly, while the input area remains fixed, making clutching more of an issue. For example, laptops are available with 38cm (15") displays with resolutions in excess of 1400 × 1050 pixels, yet the touch pad input space remains at about 4cm. With wall-sized displays, the difference is even greater.

Clutching can be reduced by increasing the ratio of display movement to control movement (*Control-Display gain*, or *CD gain*), but high CD gain can hurt performance [1,12,13,26]. An alternative is to dynamically adjust CD gain based on the input velocity. Called *pointer acceleration*, [12,21] this technique uses low CD gain at low velocity to improve precision and high CD gain at high velocity to cover large distances with minimal clutching.

Clutching can be avoided altogether by using a rate control device such as the TrackPoint [26]. This may increase performance for long distance movements, but for shorter movements, where a position control device could be used without clutching, performance will suffer [9].

To preserve the benefits of medium-distance position control and still accommodate long movements without clutching, simple hybrid position-and-rate control techniques have been proposed [2,22]. But without any haptic feedback, the transition between position and rate mode is difficult to distinguish and the rate is difficult to control. Zhai found that elastic feedback is well suited for rate control [26] and Dominjon et al. used elastic feedback for 3D hybrid position-and-rate control [8]. However, their mapping function has trajectory and velocity discontinuities when transitioning from position to rate control, further highlighting the challenges in designing a usable hybrid device.

In this paper we present RubberEdge, a 2D hybrid position-and-rate control technique using elastic feedback. Unlike past work, we designed a mapping function which enables a smooth transition from position to rate control. We conducted an experiment to evaluate its performance and explore the interaction of CD gain and pointer acceleration. We found that our hybrid control technique outperforms position-only control by 20% with a small input area similar to a laptop touch pad. We derive two predictive models for selection time with clutching and hybrid control. Finally, we discuss a class of RubberEdge devices (Figure 1) and present our first physical RubberEdge prototype device for laptop touch pads, with initial user feedback.

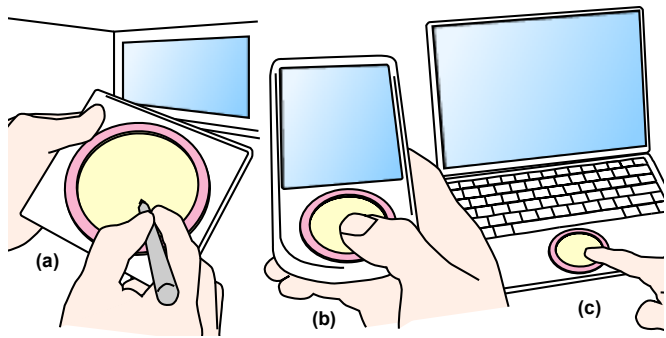

Figure 1: Design Concepts for RubberEdge Devices: (a) handheld pen tablet for a large display; (b) PDA with touch pad; (c) laptop touch pad

## BACKGROUND AND RELATED WORK
### Isotonic Position and Elastic Rate Control

Zhai defines three classes of devices, *isotonic*, *isometric* and *elastic* [26]. Isotonic devices are free-moving and use position for input. Isometric devices do not perceptibly move and use force for input. In between isotonic and isometric devices are elastic devices, where resistance increases with displacement. Elastic devices can use either position or force for input.

With any class of input device, there are different ways to map to output, the two most popular being *position control* (zero-order) and *rate control* (first-order) [26]. Zhai found that isotonic devices are better suited to position control whereas isometric and elastic devices should use rate control [26]. The latter require a self-centering mechanism – a way for the device to return to a neutral rate control state – difficult to achieve with isotonic devices. Rate control maps input to a velocity vector and moves the display pointer in that direction and speed. Position control maps device input to an output position. *Absolute* position control assigns a unique output position to every input position [5,7]. This is typically done when the input and output spaces are coincident, like on a pen-input display. *Relative* position control uses a displacement vector from an *initial input position* to the current input position [5,7]. The current output position is calculated with the vector and a corresponding initial output position. The definition of this initial input position is achieved by *clutching,* where the input stream is suspended as the device is repositioned at a new initial input position.

A transfer function uses *Control Display Gain* (*CD gain*) to scale the relative displacement vector. Jellinek et al. [12], Accot et al. [1] and Zhai [26] found that performance degraded at low and high CD gain levels. This may be partially due to limits of human motor control [3] and limited device resolution preventing selection of every pixel at high levels, and increased clutching at low levels [12, 16]. Unfortunately the effect of CD gain is not conclusive [1,12,13,15].

The amount of clutching is dependent on the maximum area of unconstrained physical movement (the *operating range*[1]), target distance, and transfer function. A small operating range causes more clutching with large target distances, and the maximum operating range is dependent on the transfer function. High CD gain increases the size of the operating range but with a potential performance cost. In theory, *Pointer Acceleration* [21]*,* a dynamic transfer function which uses high CD gain for long, fast movements and low CD gain for slow precise movements, should help minimize clutching without the performance cost of a uniform high CD gain. However, there is no published research showing a benefit for pointer acceleration [12, 19], and its effect on clutching has yet to be been shown.

---

[1] Jellinek & Card [12] use the term *footprint*, but this can be confused with the static area occupied by an object.

### Hybrid Position-Rate Control

Hybrid position-rate control techniques combine both input control modes into one device. This can be done *simultaneously* with two different physical position and rate controls mapping each to different outputs [28]; however, controlling just the pointer position with two controls simultaneously is not feasible. A more general solution is to make the device bimodal, using either position or rate control, and always controlling the pointer position directly. A common example is used in many common applications utilizing scrolling windows. When dragging and selecting items, the input switches from position to rate control as the pointer crosses the boundary of the visible window. In practice, it is difficult to move in arbitrary 2D directions and rate control is difficult, because without feedback the position-to-rate transition point is difficult to perceive and self-centre [26].

In virtual environments, Bowman and Hodges' Stretch Go-Go technique [2] uses visual feedback to help control the rate and self-centre. A virtual hand is controlled with position control, but the arm length is expanded or contracted with (constant) rate control when the hand enters circular near or far regions. The use of circular zones allows rate-control movement in any direction. Tactile 3D [24] is a commercial 3D file browser using hybrid position-to-rate control with visual and audio feedback. Rate control is used to rotate the camera with rotation speed proportional to the distance from the circular zone.

Synaptics touch pads include a hybrid technique called EdgeMotion™ [22]. At the edge of the touch pad, an isometric rate control mode is activated by switching to a downward pressure. In practice, transitioning from horizontal movement to vertical pressure for rate control may not be intuitive. Also, because of the rectangular shape of the position control zone, continuing pointer movement in the same direction in the rate control mode is difficult. No user evaluations have been reported.

Dominjon et al.'s 3D hybrid position-rate control technique uses elastic feedback with a large Virtuose 6 DOF force feedback device [8]. A spherical volume is simulated in physical space and visualized as a transparent sphere on the display. When the input point is inside the volume, movement is by position control with constant CD gain. When the input is moved beyond the spherical volume, the device uses rate control with elastic feedback. However, when we adapted their straightforward mapping functions to 2D, it exposed trajectory and velocity discontinuities at the transition point affecting its usability. Moreover, to our knowledge the authors have not conducted any sound user evaluation, and there is no satisfactory theoretical basis.

No previous examples of hybrid position-to-rate control devices have demonstrated a benefit, and little work has been done on the effect of clutching. We present our experiment which investigates these related issues later.

## RUBBEREDGE HYBRID CONTROL

Through an analysis of Dominjon et al.'s [8] straightforward mapping functions, we were able to determine the reason for erratic behaviour when transitioning to rate control. This motivated our design for improved Rubber-Edge mapping functions with a smooth transition from isotonic position control to elastic rate control.

### Straightforward Mapping Functions

Dominjon et al.'s [8] mappings (Equations 1, 2) introduce trajectory and speed discontinuities when transitioning from isotonic to elastic zones. In Equation 1, the feedback force $F$ is proportional to the distance between end effector $P$ and the isotonic-to-elastic boundary $N$ given spring stiffness $k$. The force direction is always radial with $\vec{r}$, the radial direction from the centre of the isotonic circle to P. In Equation 2, the input control rate $V$ is a third degree polynomial with a scaling constant $K$. Dominjon et al.'s implementation set $k = 200\ N.m^{-1}$ and $K = 0.03\ N^{-3}.s^{-1}$.

$$\vec{F} = -k \cdot (P - N) \cdot \vec{r} \quad (1)$$
$$\vec{V} = K \cdot F^3 \cdot \vec{r} \quad (2)$$

This formulation introduces a trajectory discontinuity as long as the isotonic trajectory is not radial to the isotonic circle. The pointer will jump to the radial trajectory defined by Equation 2 the moment it enters the rate control zone, regardless of its initial path (Figure 2). A speed discontinuity also occurs because according to equation 1, the initial force in the elastic zone will be zero, and thus the velocity will be set to zero with equation 2. Continuity of speed is important, since a noticeable drop could affect the pre-planned trajectory, impairing user performance [20].

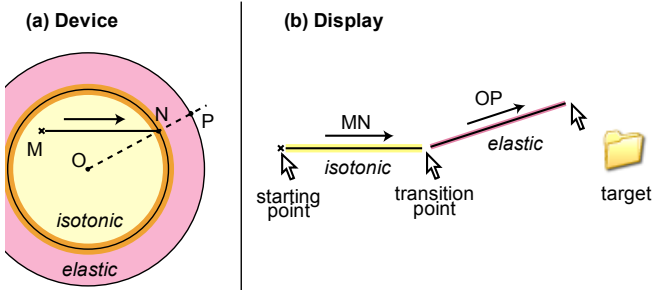

Figure 2: Trajectory Discontinuity with Straightforward Mapping Function: (a) Using the device to select a distant target, the user moves from position *M* to *N* in the isotonic zone, then transitions to the elastic zone; (b) On the display, the pointer will deviate from its trajectory of *MN* at the transition point, instantly changing to *OP* because the elastic zone always uses a direction vector radial from *O* through transition point *N* to an end effector *P*.

### RubberEdge Mapping Functions

Our mapping functions enable a consistent trajectory by rotating and translating the isotonic-to-elastic boundary after transition and we smooth the transition velocity by mixing pre- and post-transition velocities. At first it appears that simply using the same isotonic direction vector (*MN* in Figure 2) as the direction vector $\vec{r}$ for the rate in Equation 2 is the solution. However, the pre-planned trajectory direction in the isotonic zone is not always correct and the user may want to adjust it in the elastic zone. This could be done by saving the exit point $N$ and translating the pointer according to *NP*. However any change in $P$ produces an important variation in the pointer direction.

To create a consistent trajectory we translate and rotate the isotonic-to-elastic boundary zone as the user penetrates the elastic zone. We do this smoothly, by giving mass and inertia to the boundary zone using a simple physical simulation to align it with the isotonic direction vector. The intuition behind this technique is to consider how a real circular object, like a dinner plate, would rotate and translate when pulled by a string attached to its edge (Figure 3). When the user exits the isotonic zone, the exit point $N$ is saved. In the elastic zone, the vector from $N$ to the end-effector $P$ gives the force direction applied by the user on the plate. By applying angular momentum, $N$ rotates smoothly to $N'$ and the force direction vector becomes radial to $O$, the centre of the isotonic-to-elastic boundary. Past user interface researchers have utilized similar physics-based rotation and translation functions, but for rotating graphical objects with direct manipulation [14] and smoothly rotating or peeling back GUI windows [4].

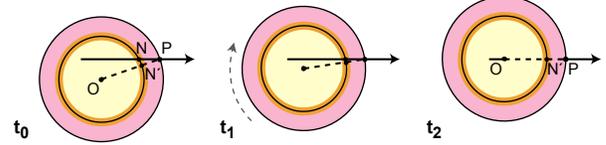

Figure 3: Rotation and translation with momentum over time: like pulling a dinner plate with a string.

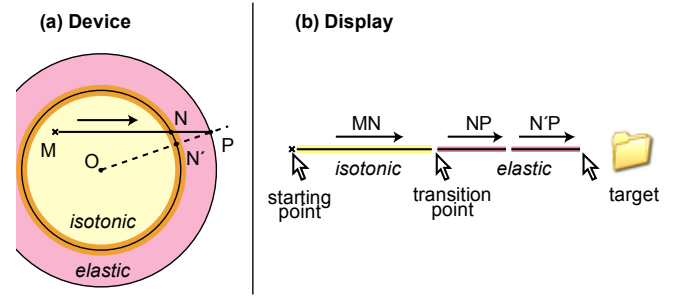

Figure 4: Continuous Trajectory with Enhanced Technique: (a) Using the device, the user moves from *M* to *N* in the isotonic zone, then transitions; (b) On the display, the initial trajectory *NP* smoothly changes to *N'P* by applying angular momentum.

The angular speed of the isotonic zone is computed by the theorem of angular momentum (Equation 3). $\omega$ is the rotation vector of the isotonic zone and $J$ is its moment of inertia with a friction term ($\mu$) added to avoid instability. The translation is proportional to the vector *NP*. We found that using a mass of $1Kg$ and a friction coefficient of $3*10^{-3}\ N.s.rad^{-1}$ smoothes out the trajectory nicely without the sharp direction changes.

$$J \frac{d\vec{\omega}}{dt} - \mu\ \vec{\omega} = \overrightarrow{ON} \wedge \overrightarrow{NP} \quad (3)$$

To smooth the transition velocity, we mix the pre-transition velocity in the isotonic zone $V_0$ with the input control rate computed in the elastic zone. Equation 4 find the mixed velocity $V_t$ where $t$ is the time after the isotonic exit and $A$ is a constant to adjust the mixing time ($A=0.3s$).

$$V_t = \left(V_0 \cdot e^{-At} + K \cdot NP \cdot (1-e^{-At})\right)\frac{\overrightarrow{NP}}{NP} \quad (4)$$

While iterating this design we found the technique worked well except when the displacement is tangent to the boundary circle. Then, the elastic force is near zero, making control difficult. However, the friction coefficient makes this occurrence rare.

## EXPERIMENT

Since no previous work has demonstrated a benefit for hybrid position-to-rate control, we conducted an experiment comparing RubberEdge to pure position control. Note that after our theoretical analysis established that the 2D adaptation of Dominjon et al.'s unproven technique has control discontinuities, it cannot be considered state-of-the-art and an empirical comparison would be of limited value.

### Goals and Hypotheses

Since the motivation for hybrid position-to-rate control is to eliminate (or at least reduce) clutching, clutching is a key factor in the experiment. We experimentally manipulated clutching by holding the device operating range constant and adjusting the target distance. Since the transfer function affects the device operating range, we included conditions for both constant CD Gain and pointer acceleration. Pointer acceleration should reduce clutching, and could negate the benefit of hybrid control. Unlike past work [12], we use the more aggressive Windows XP/Vista pointer acceleration function [21].

*H1: The hybrid technique will outperform pure position control when there is clutching.* Clutching with a position control device takes time because the pointer movement stops as the user recalibrates their position, whereas with the hybrid technique the pointer continues to move in the direction of the target. For long distances, we expect the inclusion of an elastic zone to be an advantage for the hybrid device in spite of the lower performance of pure elastic devices. This is because hybrid control still enables isometric control for fine adjustment near the targets.

*H2: Pointer acceleration improves position control performance by reducing clutching.* A dynamic transfer function uses high CD gains at high speeds which should increase the effective operating range and reduce the amount of clutching, but without hurting low speed precision.

### Apparatus

To avoid device related confounding factors in the experiment, we simulated both 2D position control and 2D hybrid control on a Phantom Omni haptic device. The Phantom uses a stylus connected to a force-feedback armature to produce haptic feedback. By simulating both techniques with a single device, we were able to compare them without introducing extraneous intra-device differences such as ergonomics, size and sensitivity. The Phantom also enabled rapid prototyping – we could iterate the RubberEdge technique and parameters with synthesized haptic feedback.

To ensure that position control performance is not adversely affected when using the Phantom, we conducted a 4 participant pilot experiment comparing it to the mouse. We used 3 target distances (70, 140, 280 mm) and a constant CD gain of 2. With these settings, no clutching was needed by either device (we found constraining the maximum mouse operating range difficult, so did not compare it with clutching). A keyboard key was used for target selection, since the Phantom and mouse have different buttons. We found mean movement times of 1.24s for the Phantom and 1.23s for the mouse. Fitts' Law analysis gave similar regression coefficients: T = -0.03 + 0.24 ID for the Phantom ($R^2=0.97$) and T = 0.09 + 0.22 ID for the mouse ($R^2=0.99$). This is consistent with previous mouse results [17]. Although not definitive, the results of this pilot bolstered our confidence that using the Phantom would be comparing position control comparable to a mouse, perhaps the best performing position control device.

### *Simulating the Techniques on the Phantom*

For both techniques, the Phantom stylus moves on a simulated haptic surface 1cm above the desk. The size of the isotonic area in each technique was constrained to a circle 40 mm in diameter by simulating a vertical wall around the perimeter (Figure 5). This size was selected to be similar to typical laptop touch pads. For the hybrid control technique, the elastic zone was accessed beyond the perimeter wall by pushing against a simulated radial spring with a stiffness of 60 N.m$^{-1}$. This setting was chosen after running a pilot experiment testing different stiffness values and it approximates the elasticity of a typical thick rubber band. For the position control technique, the simulated perimeter wall was rigid, and clutching was performed by lifting the pen above the simulated surface. To avoid instability when selecting a target with the button on the Phantom stylus, participants instead pressed a keyboard key with their non-dominant hand.

Our experiment was conducted on a 3 GHz PC with dual 19 inch, 85 DPI LCD monitors. Our C++ software displays the stimulus at 60 Hz. The Phantom Omni has a 450 DPI nominal resolution with 1000Hz haptic rendering.

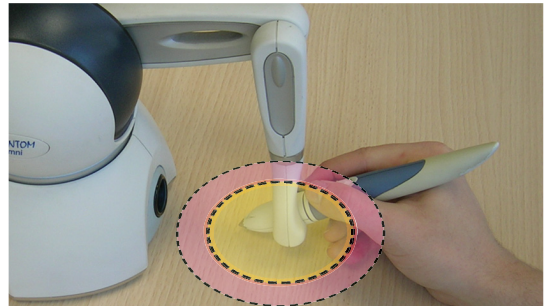

Figure 5: Simulating the RubberEdge hybrid technique with the Phantom haptic device. A 2D simulated haptic surface constrains the pen movement and the elastic zone is created using force feedback.

## Task and Stimuli

The task was a reciprocal two dimensional pointing task, requiring participants to select round targets back and forth in succession. The positions of the targets were randomly pre-computed using the position of the previous target and the current distance. When participants correctly selected a target, the target disappeared and the next one appeared on the other side of the screen. If a participant missed a target, a sound was heard and an error was logged. Participants had to successfully select the current target before moving to the next one, even if it required multiple attempts. This prevented participants from "racing through the experiment" by clicking anywhere. To avoid using the edges to assist in target acquisition, the pointer was not constrained to the bounds of the screen. Participants were encouraged to take breaks between sets of trials.

## Participants

Eight people (5 male, 3 female) participated with a mean age of 26.3 (SD = 1.5). Three participants used Windows XP/Vista pointer acceleration exclusively, two did not, and the remaining used both.

## Design

A repeated measures within-subjects design was used. The independent variables were *Technique* (*Position* control and *Hybrid* control), *Transfer Function* (*CG* - constant gain and *PA* - pointer acceleration), target *Distance* ($D_L$ – 688mm, $D_M$ – 344mm, $D_S$ – 172mm), and target *Width* ( $W_L$ – 8mm, $W_M$ – 4mm, $W_S$ – 2mm). The nine *Distance-Width* combinations give five Fitts' indices of difficulty (ID) [18] ranging from 4.5 to 8.4. We selected long distances to promote clutching, so our ID range is high. With short distances (and corresponding low IDs) the *Hybrid* and *Position* control techniques are equivalent since the elastic zone is not needed in the *Hybrid* technique.

For the *CG Transfer Function* we used a constant CD gain of 2 to encourage clutching. For the *PA Transfer Function*, we used the default Windows XP/Vista setting [21]. Using this setting, the CD gain increases continuously with the speed, from about 1.6 for low speeds to 7.3 for high speeds.

The presentation order of the 2 *Techniques* and 2 *Transfer Functions* was fully counterbalanced across participants. For each *Technique* and *Transfer Function* combination, participants completed a training period of approximately 5 minutes. Each *Distance-Width* combination was repeated 36 times with 4 Blocks of 9 trials each. *Distance-Width* combinations were presented in ascending order of ID within a single block allowing participants to leverage repetitive, ballistic movements while steadily increasing task difficulty.

After all blocks were completed for a *Technique* and *Transfer Function* combination, a short questionnaire asked participants to compare it to the previous combination. At the end of the experiment, a final questionnaire asked for an overall ranking of the four *Technique* and *Transfer Function* combinations. The experiment lasted approximately 120 minutes.

In brief, the experimental design was:

8 *Participants* × 2 *Techniques* × 2 *Transfer Functions* × 4 *Blocks* × 3 *Distances* × 3 *Widths* × 9 repetitions = 10,368 total trials.

## RESULTS

The dependent variables were movement time, error rate, and measurement of clutching and elastic zone usage.

### Error rate

Participants had an overall mean error rate of 1.5%, and a repeated measures analysis showed no significant effect of the different independent variables on error rate. In this type of experiment, a 4% error rate represents a good trade combination of speed and accuracy, so our lower error rate suggests greater emphasis on accuracy. As a result, movement times were somewhat higher and we computed the effective width for our Fitts' Law analysis [18].

### Selection Time

Selection time is the time from the beginning of the trial until the first target selected attempt. Targets that were not selected on the first attempt were marked as errors, but were still included in the timing analysis (the analysis was run with and without error trials and the same significant effects were found with similar F and p values).

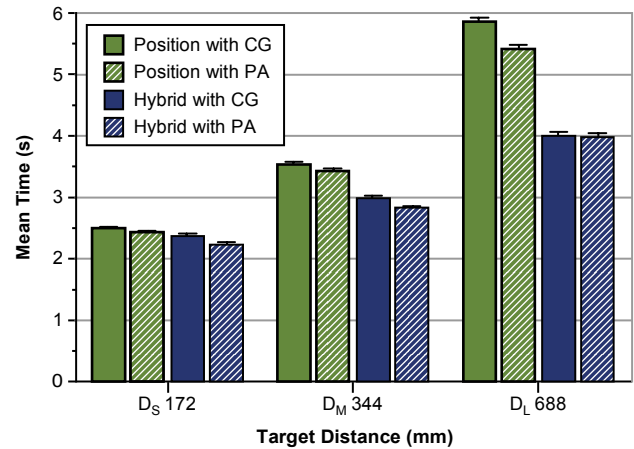

Figure 6: *Technique* × *Transfer Function* × *Distance* interaction on selection time (error bars 95% CI).

Repeated measures analysis of variance showed that the presentation order of *Technique* and *Transfer Function* had no significant effect on movement time, indicating that a within-participants design was appropriate. No significant effect for *Block* was found indicating that there was no learning effect present. There was a significant main effect for *Technique* on selection time ($F_{1,7}$ = 16.0, p < 0.005 ) with the *Hybrid* control technique outperforming the *Position* control technique by 20.6%. As expected in a target selection experiment, there were also significant main effects for *Distance* ($F_{2,14}$ = 471.0, p < 0.0001) and *Width* ($F_{2,14}$ = 231.0, p < 0.0001) on selection time. The significant interactions for *Technique* × *Distance* ($F_{2,14}$ = 50.2, p < 0.0001) and *Technique* × *Transfer-Function* × *Distance*

($F_{2,14}$ = 9.7, p < 0.017) are perhaps most relevant. These show that the selection time increases with *Distance* at different rates given the *Transfer-Function* (Figure 6). Pair-wise comparisons found no significant difference between the two *Techniques* for the smallest *Distance* $D_S$ but significant differences for $D_M$ (p < 0.017) and $D_L$ (p < 0.001) with 16% and 29% improvements for *Hybrid* control over *Position* control respectively. The high selection times for distant targets with the *Position* control technique are due to heavy clutching, which we discuss in detail below.

Pair-wise comparisons revealed a significant difference between the two *Transfer Functions* for the *Position* technique and $D_L$ (p < 0.032). *PA* reduces the selection time by 7.5% compared to *CG*. It appears that participants were able to harness the higher speeds for distant targets and thus use higher CD gains to avoid clutching.

**Fitts' Law Analysis**
The significant interaction between *Technique*, *Transfer-Function* and *Distance* leads us to a Fitts' Law analysis. We aggregated the *Distance-Width* combinations for each *Technique* and *Transfer-Function* and computed the effective width since the error rate is not equal to 4% [18].

Unlike many past studies, we found poor regression fitness suggesting that Fitts' Law may not hold in the presence of significant clutching or for a technique combining two different control mappings. The index of difficulty (ID) is expressed as a ratio between target distance and width, giving the same importance to each. Therefore, Fitts' Law predicts that any *Distance-Width* combination with the same ID will yield the same selection time. However, in looking at a plot of ID and selection time (Figure 7), it appears that distance alone affects position control clutching or encourages hybrid control mode switching.

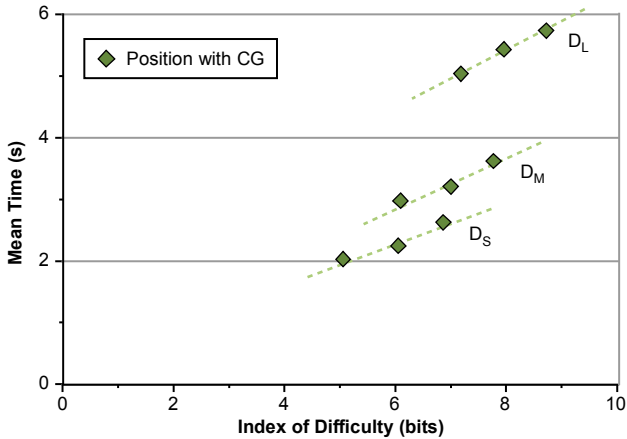

Figure 7: Selection times given *Distance-Width* combinations for the *Position* technique and *Constant* transfer function. Trend lines for *Distance-Width* groups are highlighted, suggesting *Distance* has a greater effect on selection time. Plots of other *Technique* and *Transfer Function* combinations are similar.

**Usage of Clutching and Elastic Zone**
To help characterize technique usage and to investigate possible explanations for the poor conformance to Fitts' Law, we analyzed the amount of time spent clutching in *Position* or using the elastic zone control in *Hybrid*, as well as the number of invocations for each. Since cross-technique statistical comparisons of these two measures would not be meaningful, separate ANOVAs were used for each technique.

| Technique | a | b | $r^2$ |
|---|---|---|---|
| *Position* with *CG* | -3.9 | 1.1 | .73 |
| *Position* with *PA* | -2.4 | 0.9 | .70 |
| *Hybrid* with *CG* | -1.0 | 0.6 | .86 |
| *Hybrid* with *PA* | -1.0 | 0.6 | .74 |

Table 1: Fitts' Law regression values for *Technique* and *Transfer Function*: *a* is the intercept of the regression line, *b* is the slope, $r^2$ is the fitness.

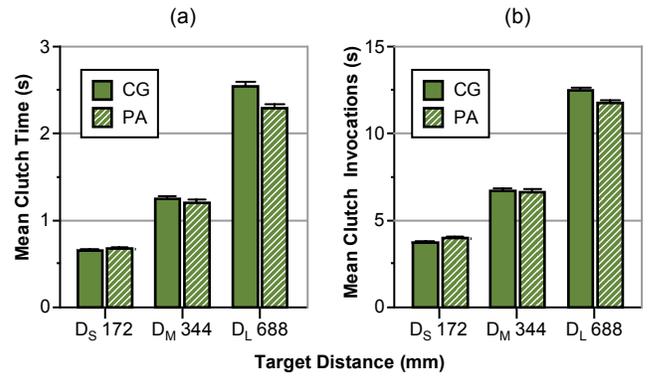

Figure 8: Position Control Transfer Func. × Distance interaction on: (a) clutch time; (b) clutch invocations

*Clutch Time and Elastic Zone Time*
Clutch time is the total time in which the stylus was lifted during a trial, and elastic zone time is the time spent outside the isotonic zone. For the *Position* technique, there is a significant main effect of *Distance* ($F_{2,14}$ = 165.5, p < 0.0001) showing that clutch time increases as *Distance* increases. Pair-wise comparisons revealed that the total clutch time increased from 0.7s for $D_S$ to 1.2s for $D_M$ and 2.4s for $D_L$ for *Position* control and *CG* (all p < 0.0001). Results for *PA* are similar (Figure 8a). A *Transfer Function* × *Distance* interaction ($F_{2,14}$ = 21.9, p < 0.0001) and pair-wise comparison show that clutch time is reduced by 10% with the *PA* Transfer function for the largest distance $D_L$ (p = 0.018). For the *Hybrid* technique, we found a significant main effect for *Distance* ($F_{2,14}$ = 956.7, p < 0.0001) showing that elastic zone time increases with *Distance*.

*Clutch and Elastic Zone Invocations*
Comparing the mean number of *Position* invocations to *Hybrid* elastic zone transitions can help characterize technique usage. With 1.3 and 7.5 invocations respectively, we see that frequent clutching actions can be replaced often with a single transition to the elastic zone.

For the *Position* technique, a significant main effect was found for *Distance* ($F_{2,14}$ = 430.6, p < 0.0001), but the *Transfer Function* × *Distance* interaction ($F_{2,14}$ = 19.1, p < 0.0001) is more relevant. Pair-wise comparison revealed that the number of *Position* clutch invocations for $D_L$ is dependent on *Transfer Function*: 12.4 for CG and 11.6 for PA (p = 0.013) (Figure 8b). *Hybrid* invocations were near 1 regardless of *Distance* or *Transfer Functions*.

Table 2 compares the actual clutching invocations with an ideal minimum number calculated by dividing the floor of target distance by maximum device operating range (accounting for CD gain). We can see that the ratio remains constant around 0.75, indicating that participants did not use the maximum available operating range for each clutch.

|  | *Position* with *CG* | | |
|---|---|---|---|
| Distance | $D_S$ | $D_M$ | $D_L$ |
| Ideal | 3 | 5 | 9 |
| Actual | 3.75 | 6.7 | 12.5 |
| Ratio | 0.8 | 0.75 | 0.72 |

Table 2: Comparison of Actual and Ideal numbers of invocations used in the Position technique.

### User Feedback

Overall preference for *Technique* was split. Those preferring *Hybrid* found it faster for long-distance targets and disliked the repetitive clutching motion with *Position*. Those preferring *Position* found it acceptable to use conventional clutching for short and medium distances, and felt that it was difficult to accurately exit the elastic zone and fine-tune their selection. Our observations during the experiment reinforced this last comment: when participants exited the elastic zone having undershot the target, any movement back in the direction of the target transitioned them back to the elastic zone. This left no room for isotonic movement to fine-tune the selection. One way to address this is to allow conventional clutching in the isotonic zone of the *Hybrid* technique. Later in the paper, we present a prototype device which does exactly that.

### DISCUSSION

Our experiment confirmed our two hypotheses and illustrated a negative performance impact with clutching.

Our results confirmed hypothesis H1: a hybrid position-rate control technique has a performance advantage over pure position control when faced with significant clutching. Our experimental design intentionally provoked clutching, and overall we found *Hybrid* control improved performance by 20.6%. Specifically, *Hybrid* control improved performance by 16% and 29% over *Position* control for $D_M$ and $D_L$ respectively. Increased clutching with position control appears to be the reason. Clutch times went from 0.7s at $D_S$ to 1.2s at $D_M$ and 2.4s at $D_L$ with the *CG* transfer function. One reason why clutching may be slower than *Hybrid*, is the number of invocations required. Users had to clutch more than 12 times to reach $D_L$.

We confirmed hypothesis H2: pointer acceleration reduces clutching with *Position* control. The effect is somewhat slight; we saw it only at the longest distance where PA clutch time was 10% lower, requiring an average of 11.6 invocations compared to 12.4 for CG. This suggests that participants are able to utilize the high CD gain levels with quick ballistic movements. Our results differ from past researchers who did not see an effect [12]. We attribute this to using a more aggressive pointer acceleration function and a task requiring significant clutching.

The experiment demonstrates the negative impact of clutching on user performance and shows that selection times with significant clutching do not conform to Fitts' Law. With clutching, task difficulty appears to be primarily dependent on distance, rather than the ratio of distance to width as in Fitts' Law. Past researchers have not reported this [12, 16], perhaps because their experiments did not promote significant clutching.

### FORMAL MODELS

We developed two formal models to predict position control performance with clutching, and performance with a hybrid position-rate control technique.

### Clutching Model

We base our model for position control movement with clutching on a two-part, idealized movement. In the first part, the user clutches several times to bring the pointer within a "clutch-free" distance to the target. In the second part, the user completes the movement without clutching, and selects the target. The total movement time $T$ is the sum of the time for the first part $T_1$ and the second part $T_2$:

$$T = T_1 + T_2 \quad (5)$$

$T_1$ is dependent on the number of clutches N and the time for each clutch $T_C$, which we assume to be constant. We also assume the time the cursor is engaged *between* two clutches to be equal to $T_C$, hence the factor 2:

$$T_1 = 2 \cdot N \cdot T_C \quad (6)$$

$N$ is dependent on the target distance $D$ (in mm) and the effective device operating range $d_e$:

$$N = \left\lfloor \frac{D}{d_e} \right\rfloor \quad (7)$$

The effective operating range of the device, $d_e$ is calculated from the physical device operating range $d$ (in mm), the CD gain $CD$, and a corrective parameter $c$. Recall that our experimental results showed that in practice, only a portion of the operating range is actually used:

$$d_e = c \cdot d \cdot CD \quad (8)$$

The movement time in the second part, $T_2$, can be calculated using Fitts' Law [18] with the remaining target distance $D_2$ and Fitts' device parameters $a_i$ and $b_i$.

$$T_2 = a_i + b_i \log_2\left(\frac{D_2}{W} + 1\right) \quad (9)$$

Where $D_2$ is equal to:

$$D_2 = D - N \cdot d_e \quad (10)$$

By substituting Equations 6 to 10 into Equation 5 and simplifying, we have a model which accounts for clutching when predicting target selection time:

$$T = 2\,N\,T_c + a_i + b_i \log_2\left(\frac{D - N \cdot c \cdot d \cdot CD}{W} + 1\right) \quad (11)$$

## Hybrid Control Model

A similar idealized model exists for a hybrid position-rate control technique. Similar to clutching, the movement has two parts (Equation 5). $T_1$ is the time to move to the isotonic-elastic boundary and $T_2$ is the remaining time in the elastic zone to the target. Note that if the target is within reach of isotonic movement, then $T_2=0$ and $T_1$ can be predicted by Fitts' Law with the parameters $a_i$ and $b_i$ of an isotonic device. Otherwise, we suppose the movement distance in the first part is equal to the effective device operating range and $T_1=T_C$. $T_2$ can then determined using Fitts' Law with parameters $a_e$ and $b_e$ for an elastic device, and the remaining distance $D_2$ to the target:

$$T_2 = a_e + b_e \log_2\left(\frac{D_2}{W}+1\right) \quad (12)$$

Where D2 is simply:

$$D_2 = D - CD \cdot d \quad (13)$$

By substituting Equations 12 and 13 into Equation 5 and simplifying, we have a model for hybrid movement predicting target selection time:

$$T_a = T_c + a_e + b_e \log_2\left(\frac{D - CD \cdot d}{W}+1\right) \quad (14)$$

## Comparison to Experimental Results

To test the validity of our models, we compared their predicted selection times with the results of our experiment. The following model parameters were used *d=40mm*, *CD=2*, *c=0.75*, *$T_C$=0.2s* ($T_C$ from our experiment). The Fitts' law parameters were from the literature $a_i=0$, $b_i=4.5$ [17,9], $a_e=0$, $b_e=2.0$ [9]. Considering the simplicity of the model, we found good fitness (Figure 9). The root mean square (RMS) is 0.4s for clutching and 0.2s for hybrid (The RMS for Fitts' law are respectively 1.1s and 0.6s). At $D_S$, the clutching model was 25% lower, likely due to the floor in Equation 7, while the experimental data presents a mean value. For example, at $D_S$, the predicted number of clutches is 3, but the experimental data is 3.75.

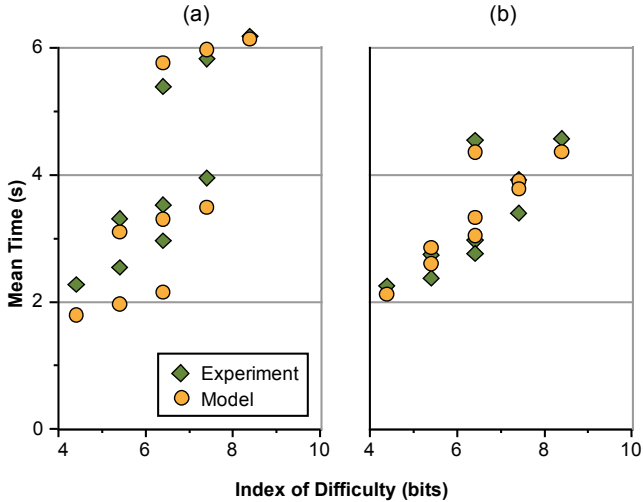

Figure 9: Predicted Model Time vs. Actual Time for: (a) Clutching Model; (b) Hybrid Model.

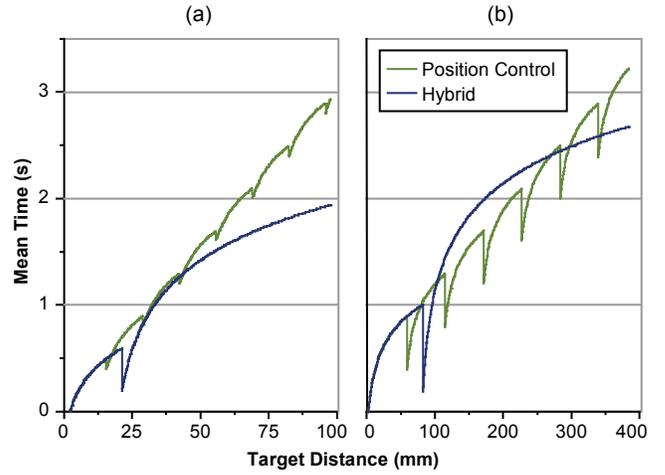

Figure 10: Theoretical Comparison for *W=4mm*: (a) Touch Pad PDA: *d=10mm, CD=2* (b) High Resolution Laptop: *d=40mm, CD=2*.

## When is Hybrid Control Advantageous?

We can use the models to predict when hybrid control will have a performance advantage over position control.

For example, consider a laptop with a 38cm (15"), 1400 × 1050 pixel display, and 4cm touchpad. Using our theoretical model, RubberEdge hybrid control will outperform position control when targets are more than 30cm apart (nearly three-quarters of the maximum possible target distance) (Figure 10b). A second example is the HP iPAQ hx4700 PDA which has a 1cm touchpad and a 10cm (4") display. Here our model predicts an advantage for hybrid control above 5 cm (half the display distance) (Figure 10a).

The examples in Figure 10 also illustrate a potential drawback with hybrid control. Depending on the operating range size, one or two manual clutches can be faster than using the rate control zone. In our model this is attributed to the lower performance of elastic devices, but there may be other factors not accounted for, such as a constant mental transition time. Regardless of the reason, it appears that a hybrid device should allow standard isotonic clutching as well as an elastic zone. This way, the user can develop their own optimized strategy for reaching near or far targets.

## PROTOTYPE DEVICE

With this more flexible hybrid model in mind, we built a device prototype that enabled a mix of isotonic clutching and elastic rate control. Our initial requirements were:

- cheap and compatible with current notebooks
- support for high resolution absolute input to measure the elastic zone penetration and compute pointer velocity
- support for relative position input and clutching in the same way as an existing device

We found that modifying a standard laptop touch pad fulfilled these requirements. Creating an elastic zone with the right feel and stiffness similar to the Phantom required some trial and error. We experimented with elastic materials like latex gloves, balloons, and elastic fabrics mounted on different types of frames. Eventually, we converged on a simple design using a 1mm thick plastic frame cut from a old phone card. The frame has a 40mm hole with a plastic ring suspended by four rubber bands for elastic feedback (Figure 11). The ring is 36mm in diameter leaving 2mm for elastic movement. Plastic lets the ring slide easily on the touch pad surface and has just enough tactile feedback to define the boundary of the 30mm isotonic zone. We would have preferred creating a larger isotonic zone, but the borders of the frame had to support the elastic force. Adding physical constraints in this way is reminiscent of Wobbrock et al's EdgeWrite [25].

Our driver uses the Synaptics SDK [23] to measure the absolute finger position at 2000 DPI. In the isotonic zone, the pointer behaves like a standard Windows touch pad. When the finger enters the elastic zone, we transition using the RubberEdge mapping functions and compute the pointer velocity (Equations 3, 4). Our driver works like a standard Windows' pointing device with any application.

Early designs revealed that isotonic-elastic boundary accuracy is critical since there is only 2mm of movement. Imperfections in our fabrication and the non-uniform way in which a finger contacts positions around the ring led us to develop a two-step calibration (Figure 12). First, the boundary is defined by tracing around the perimeter of the ring. Then, to calibrate the maximum force (penetration distance) in each direction, the user pushes into the elastic zone at eight radial positions. We interpolate between these measurements when computing elastic rate control.

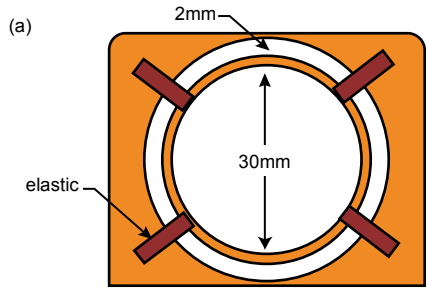

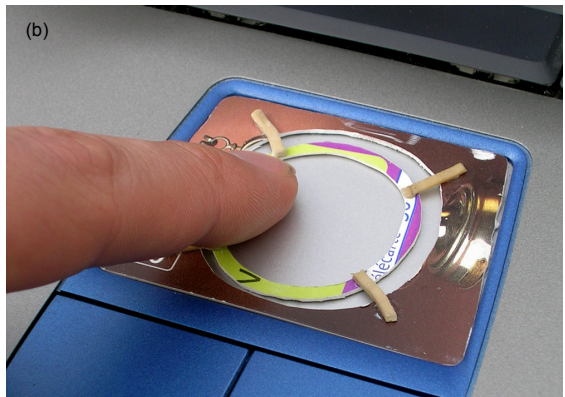

Figure 11: RubberEdge prototype device: (a) Schematic; (b) implementation.

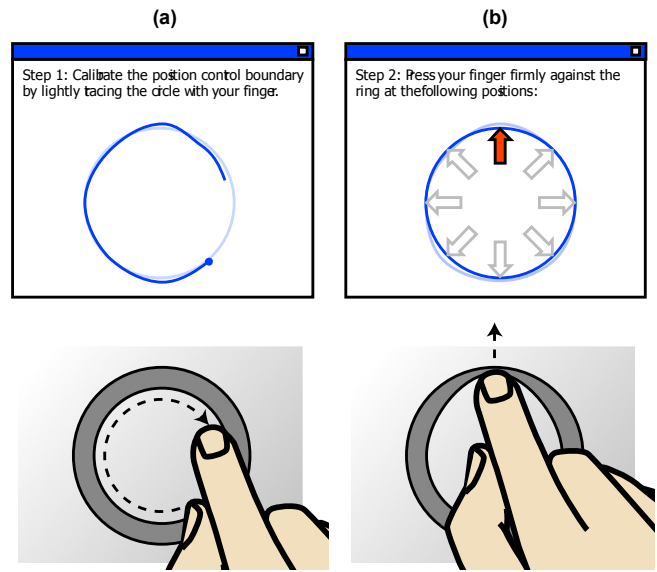

Figure 12: Two Step Prototype Calibration addresses caused by finger angle and fabrication: (a) Calibrating the boundary of the isotonic zone by tracing the finger clockwise around the perimeter; (b) Calibrating the maximum force by pushing the finger into the elastic zone at eight radial positions.

**Initial Evaluation**

We ran a pilot study with four participants to gather initial feedback about our prototype. We targeted people with touch pad experience since we were interested in the usability of our device, not touch pads in general; 3 participants used a touch pad daily and the fourth occasionally. All were right-handed. For approximately 20 minutes, participants used the device with common Windows tasks: file browsing, viewing PDF documents, painting, and web browsing. The tasks included pointing, window scrolling and steering through menus.

To grow accustomed to the device's reduced operating range, participants used only the isotonic zone with the rate control disabled for the first 2 minutes. We then enabled the elastic zone using a generic calibration, and gave no explanation or instructions. Participants immediately grasped that the pointer moved in two different ways depending whether you were pushing into the ring. The most difficulty was with elastic rate control: participants would at first overshoot the target, then sometimes overcompensate with hesitant and slow rate control. Past researchers have found that elastic rate control has a steep learning curve [26]. Two of the participants used overshooting as a kind of strategy: in the elastic zone, they shot the pointer as fast as possible past the target, then moved back to the target under isotonic control. However, after more practice, participants generally moved the pointer more accurately and with less hesitation using the elastic zone. Overall, participants said they liked using rate control for continuous movement of the pointer over far distances and appreciated the ability to use the isotonic zone for tasks like drawing.

## CONCLUSIONS AND FUTURE WORK

RubberEdge hybrid position and rate control enables users to reach distant targets without clutching, yet still maintains benefits of position control for precise movements. Our mapping functions eliminate trajectory and velocity discontinuities when transitioning from isotonic position control to elastic rate control. The results of our controlled experiment found that hybrid control outperforms pure position control by 20% when there is significant clutching. This advantage is in spite of our related finding that a pointer acceleration transfer function will decrease clutching. We present theoretical performance models for position control clutching and hybrid position rate control, enabling designers to determine when hybrid control is beneficial. Based on our experimental and theoretical investigations, we developed a RubberEdge hybrid device for laptops which revealed design considerations such as construction, material, and calibration. With promising initial user feedback, we plan to further iterate our current prototype and investigate other types of RubberEdge hybrid devices.